%% file: main.tex
\documentclass[runningheads]{llncs}
\usepackage{graphicx}
\usepackage{times}
\usepackage{todonotes}
\usepackage{xspace}
\usepackage[hidelinks]{hyperref}
\usepackage[inline]{enumitem}
\usepackage{amsmath,amsfonts,amssymb}
\usepackage{array}
\usepackage{multirow}

\newcommand{\sarscov}{SARS-CoV-2}
\def\Card#1{\ensuremath{\left\vert #1 \right\vert}}
\newcommand{\Prc}[2]{\ensuremath{\mathrm{Pr} \{#1 \!\left. \right\rvert\! #2  \}}}
\def\Pr#1{\ensuremath{\mathrm{Pr} \{#1 \}}}

\usepackage{algorithm}
\usepackage{algorithmic}

\makeatletter
\newcommand{\IWHILE}[1]{\ALC@it\algorithmicwhile\ #1\ \algorithmicdo\begin{ALC@whl}}
\newcommand{\IENDWHILE}{\end{ALC@whl}}
\newcommand{\IENDFOR}{\end{ALC@for}}
\newcommand{\IIF}[1]{\ALC@it\algorithmicif\ #1\ \algorithmicthen}

\makeatother
\newcommand{\covid}{COVID-19}

\DeclareMathOperator{\LEventually}{\mathbf{F}}

\DeclareMathOperator{\PProb}{\mathbf{P}}

\makeatletter
\renewcommand\subparagraph{\@startsection{paragraph}{4}{\z@}%
                       {-12\p@ \@plus -4\p@ \@minus -4\p@}%
                       {-0.5em \@plus -0.22em \@minus -0.1em}%
                       {\normalfont\normalsize\itshape}}
\makeatother
\usepackage[compact]{titlesec}
\usepackage[font=small,skip=3pt]{caption}


\usepackage[final]{listings}
\lstdefinelanguage{prism}{
language=C++,
morekeywords={module,endmodule,false,true,init,endinit,mdp},%
morecomment=[l]{--},
    keywordstyle=\color{blue!55}\bfseries,
    basicstyle=\sffamily\fontsize{6.0}{7.0}\selectfont,
    commentstyle=\rmfamily\color{green!60!black},
    breakatwhitespace,
    showstringspaces=false,
    breaklines=true,
    frame=lines,
    float=t!,
    captionpos=b,
}

\begin{document}
\captionsetup{belowskip=0pt}
\title{
  Verifying a stochastic model for the spread of a \sarscov{}-like infection: opportunities and limitations
  \thanks{This work is partially funded by the grant MOSES, Bando interno 2020 Università di Trento "Covid 19".}}
%
%
\author{Marco Roveri\inst{1}\orcidID{0000-0001-9483-3940}
\and Franc Ivankovic\inst{1}
\and Luigi Palopoli\inst{1}\orcidID{0000-0001-8813-8685}
\and Daniele Fontanelli\inst{1}\orcidID{0000-0002-5486-9989}
}
\authorrunning{M. Roveri et al.}
%
\institute{Department of Information Engineering and Computer Science\\
University of Trento -- Via Sommarive 9 - 38123 Povo - Trento (Italy)\\
\email{name.surname@unitn.it}\
}
\maketitle              
%

\sloppypar

\begin{abstract}
\input{abstract}
\end{abstract}
\section{Introduction}
\label{sec:intro}
\input{intro}

\section{Background}
\label{sec:bg}
\input{bg}

\section{A stochastic model for \sarscov{}-like infection's spread}
\label{sec:model}
\input{model}

\section{Experimenting with state-of-the-art stochastic model checkers}
\label{sec:exp}
\input{experimental}

\section{Related works}
\label{sec:rw}
\input{rw}

\section{Conclusions and future works}
\label{sec:conc}
\input{conclusions}

\clearpage
%
\bibliographystyle{splncs04}
\bibliography{biblio}
\clearpage
\appendix
\input{appendix}

\end{document}

%% file: abstract.tex
There is a growing interest in modeling and analyzing the spread of diseases like the \sarscov{} infection using stochastic models. These models are typically analyzed quantitatively and are not often subject to validation using formal verification approaches, nor leverage policy syntheses and analysis techniques developed in formal verification.

In this paper, we take a Markovian stochastic model for the spread of a \sarscov{}-like infection. A state of this model represents the number of subjects in different health conditions. The considered model considers the different parameters that may have an impact on the spread of the disease and exposes the various decision variables that can be used to control it. We show that the modeling of the problem within state-of-the-art model checkers is feasible and it opens several opportunities. However, there are severe limitations due to i) the espressivity of the existing stochastic model checkers on one side, and ii) the size of the resulting Markovian model even for small population sizes.


%% file: intro.tex
The recent \covid{} pandemic highlighted the importance to
develop reliable models to study, predict and control the evolution and
spread of diseases. Several analytical models have been proposed in
the
literature~\cite{bernoulli:1760,Kermack:1927,allen:2008,Brauer:2019,giordano:2020,ghezzi:1997,giordano:2016,Khanafer:2016,Yousefpour:2020}. %
All these models are deterministic and aims at capturing the disease
dynamics. These studies have been complemented with studies proposing
stochastic models, that differently from deterministic ones, allows to
derive richer set of informations like e.g. show converge to a
disease-free state even if the corresponding deterministic models
converge to an endemic equilibrium~\cite{anderson:1992}; computing the
probability of an outbreak, the distribution of the final size of a
population or the expected duration of an
epidemic~\cite{Brauer:2019,Sattenspiel:1990}; computing the
probability of transition between different state of \covid{}-affected
patients based on the age class~\cite{ZARDINI2021100530}; or
evaluating the effects of lock-down policies~\cite{Flavia20}.
Recently, the evolution of diseases has also been modeled with
stochastic models in form of Markov
Processes~\cite{cassandras2009introduction,allen:2008,DBLP:journals/corr/abs-2204-11317}.
The use of stochastic models opens for the possibility to use
Stochastic Model Checking techniques to
\begin{enumerate*}[label=\roman*)]
\item validate the model using probabilistic temporal properties of
  the model as well as compute quantitative measures of the degree of
  satisfaction of a given temporal
  property~\cite{10.1371/journal.pone.0145690};
\item evaluate the effects of a strategy on a population during the
  evolution of a disease~\cite{C1MB05060E,Chauhan2015}.
\end{enumerate*}
The work in \cite{DBLP:journals/corr/abs-2204-11317} describes a
stochastic compartmental model (the population has been broken down
into several compartments) for the spread of \covid like diseases,
with some preliminary results on the use of stochastic model checking
techniques to analyze a simplified version of the epidemic model.

In this paper we make the following contributions. First, we consider
the epidemic model presented in
\cite{DBLP:journals/corr/abs-2204-11317} and we show how to encode it
into languages suitable for being analyzed with state-of-the-art
stochastic model checkers. To this extent, we developed a C++ open
source tool that given the parameters of the epidemic model is able to
generate models in the PRISM formalism~\cite{kwiatkowska2011prism} to
be then analyzed by tools supporting that formalism (e.g. the
PRISM~\cite{kwiatkowska2011prism} and the
STORM~\cite{DBLP:journals/corr/abs-2002-07080} model checkers).
Second, we show that the encoding of the considered model in the
language accepted by model checkers is out of the espressivity
capabilities of the input languages, and even for small population
sizes it results in very large files that easily reach unacceptable
timings for the storage and parsing of such models, thus preventing
any further analysis. To this extent, we modified the developed tool
to link with the STORM model checker to pass the model directly in
memory without the use of intermediate files.
Third, we used the developed tool to study the model with increasing
population sizes, analyzing the models against given temporal
properties, and evaluating the effects of different control
policies. These results show that the approach is feasible, but they
confirm the scalability issues first noticed
in~\cite{nasir2020epidemics,Hak18}, and pose challenges to the
community to address large population sizes on one hand, and
espressivity requirements on the input languages, on the other hand, to
facilitate the specification of such complex mathematical models.

This paper is organized as follows. In Section~\ref{sec:bg} we briefly
summarize the basic concepts. In Section~\ref{sec:model} we discuss
the model presented in~\cite{DBLP:journals/corr/abs-2204-11317} and we show how to compute
the probabilistic transition function. In Section~\ref{sec:exp} we
describe the tools and the experiments carried out. In
Section~\ref{sec:rw} we discuss the related works, an finally in
Section~\ref{sec:conc} we draw conclusions and discuss possible future
works.

%% file: bg.tex
An Markov Decision Process (MDP) is a tuple
$\langle S, S_I, A, T, R\rangle$ where $S$ is a finite set of states,
$S_I\subseteq S$ is the set of initial states, $A$ is a finite set of
actions (i.e. control variables),
$T: S \rightarrow 2^{A \times S \times \mathbb{R}}$ is the transition
probability function that associates to a state $s \in S$, an action
$a\in A$ the probability $p$ to end up in state $s'$,
$R: S \times A \rightarrow \mathbb{R}$ is the reward function, giving
the expected immediate reward $r$ gained by for taking action
$a \in A$ in state $s\in S$ (we remark that, in many cases there is no
reward function).
A Discrete Time Markov Chain (DTMC) is an MDP such that in each state
$s \in S$ there is only one action to be considered with an associated
probability to end-up in a state $s' \in S$ (i.e. there is a single
probability distribution over successor states).
Partially Observable MDPs, extend MDPs by a set of observations and
label every state with one of these observations. Thus, the states
labeled by the same observation must be considered undistinguishable.

Several formalism have been proposed to specify (PO)MDPs and DTMCs. We
refer to~\cite{DBLP:journals/corr/abs-2002-07080} for a thorough
overview. In the following we briefly describe the PRISM
language~\cite{kwiatkowska2011prism} supported by the PRISM and STORM
stochastic model checkers. The PRISM language is a simple state-based
language such that
\begin{enumerate*}[label=\roman*)]
\item the user specifies variables with a finite domain (a complete
  assignment of a value to these variables at any given time
  constitutes a possible state of the system);
\item the behavior is specified through commands of the form
  \texttt{[action] guard -> prob\_1 : update\_1 + ... + prob\_n :
    update\_n} where: \texttt{guard} is a predicate over all the
  variables in the model, each \texttt{update\_i} describes a
  transition which the model can make if the \texttt{guard} is true (a
  transition specifies the new values of the variables, and is
  associated to the probability/rate \texttt{prob\_i} to take that
  update), and to an optional annotation \texttt{action} (modeling a
  control variable).
\end{enumerate*}
On a (PO)MDP/DTMC model one can check several kind of properties, like
e.g.,  temporal logic formulas based on PCTL~\cite{HJ94} (e.g.,
property $\PProb_{< 0.25} [ \LEventually O_k = C ]$ means the
probability of reaching a state where the variable $O_k$ is equal to
$C$ is less than 0.25), or compute the probability with which a system
reaches a certain state (e.g., $\PProb_{=?} [ \LEventually O_k = C ]$
to compute the probability to reach a state where $O_k$ is equal to $C$),
or perform conditional probability and cost queries, or compute long-run
average values (also known as steady-state or mean payoff values), or synthesize a policy to satisfy a certain PCTL property. We
refer the reader
to~\cite{HJ94,kwiatkowska2011prism,DBLP:journals/corr/abs-2002-07080}
for a thorough discussion of possible queries.

In the following, we denote with $n! = n \cdot (n-1) \cdot ... \cdot 1$ the
factorial (i.e. the permutations of $n$ elements), with
$\binom{n}{k} = \frac{n!}{k! \cdot (n-k)!}$ the binomial coefficient,
with
$\mathbb{M}_{n,n_1,n_2,\dots,n_{k-1}} = \binom{n}{n_1,n_2,\dots,n_k} =
\frac{n!}{\prod_{i=1}^k n_i!}$ the multinomial coefficient (i.e. the
permutations with repetitions obtained computing all the permutations
of $n$ elements taken from $k$ sets with $n_1, n_2,\dots,n_k$ elements
such that $n_k = n - \sum_{i=1}^{k-1} n_i$), and with
$\mathcal{B}(N, p)_X = \binom{N}{X}p^X(1-p)^{N-X}$ the binomial
probability distribution function where $X$ is the total number of
successes, $p$ is the probability of success on an individual trial,
and $N$ is the number of trials.


%% file: model.tex
\begin{figure}[t!]
  \centering
  \includegraphics[width=0.7\textwidth]{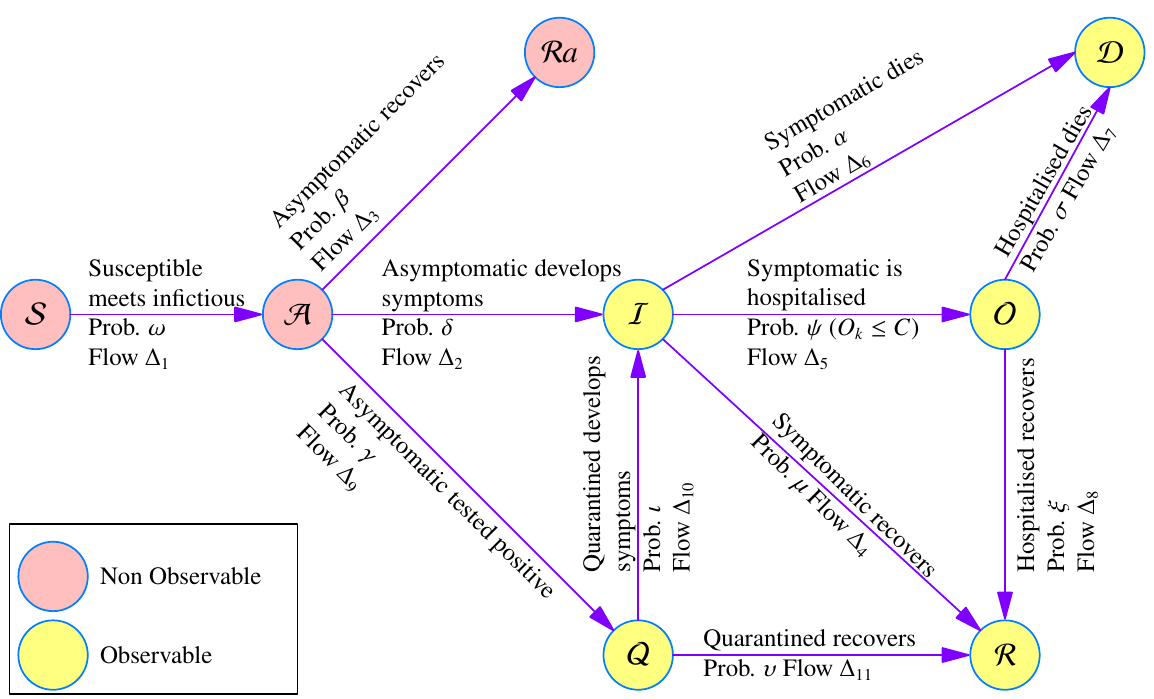}
  \caption{\label{fig:model}Transitions between the
    different states of a single subject of the entire population.}
\end{figure}
We model a subject of the population as a stochastic
discrete--time system with 8 states, as illustrated in
Figure~\ref{fig:model}, each representing a possible state of the
subject: the susceptible $S$, infected $I$, recovered $R$,
asymptomatic $A$ (i.e., a group of infected people that do not exhibit
symptoms but are infective), hospitalised $O$, dead $D$, recovered
$Ra$ from an asymptomatic state, and the case of swab-tested people
that are quarantined (denoted with $Q$) if they result positive.
The evolution is observed at discrete time $k$ and each subject can
belong to one of eight possible states.
The subjects who are in a state at step $k$ will be denoted by a
calligraphic letter (e.g., $\mathcal{S}_k$ is the set of susceptible
subjects).
Figure~\ref{fig:sym} reports the symbols used to denote the different
sets, their cardinality (e.g., $S_k$ is the cardinality of
$\mathcal{S}_k$) and the different probabilities governing the
transition of a subject between the different sets.
The states of the discrete--time Markov chain can be characterized by
a vector
$ \vec{V}_k = \left[S_k, A_k, I_k, R_k, O_k, D_k, Q_k, Ra_k \right]$
such that the values of all the different quantities are non-negative
integers representing the cardinality of their respective sets.

\begin{figure}[t!]
\centering
\bgroup
\newcolumntype{C}{>{$}l<{$}}  
\newcolumntype{L}{@{}l}
\setlength\tabcolsep{5pt}
\renewcommand{\arraystretch}{0.98}
\scalebox{0.715}{
\fbox{\begin{tabular}{CCCC}
  \multicolumn{2}{L}{\textbf{Sets}}\\
   S_k  = \Card{\mathcal{S}_k}  & \text{N. of susceptible sub. $\mathcal{S}_k$ at step } k, &
   A_k  = \Card{\mathcal{A}_k}  & \text{N. of asymptomatic sub. $\mathcal{A}_k$ at step } k,\\
   I_k  = \Card{\mathcal{I}_k}  & \text{N. of symptomatic sub. $\mathcal{I}_k$ at step } k, &
   R_k  = \Card{\mathcal{R}_k}  & \text{N. of recovered sub. $\mathcal{R}_k$ at step } k,\\
   Ra_k = \Card{\mathcal{R}a_k} & \text{N. of asympt. recovered sub. $\mathcal{R}a_k$ at step } k,&
   O_k  = \Card{\mathcal{O}_k}  & \text{N. of hospitalised sub. $\mathcal{O}_k$ at step } k,\\
   D_k  = \Card{\mathcal{D}_k}  & \text{N. of deceased sub. $\mathcal{D}_k$ at step } k, &
   Q_k  = \Card{\mathcal{Q}_k}  & \text{N. of quarantined sub. $\mathcal{Q}_k$ at step } k,\\
   Q^{(R)}_k \!\!=\! \Card{\mathcal{Q}^{(R)}_k\!}\! & \text{N. of quarantined sub. recovered $\mathcal{Q}^{(R)}_k$ at step } k.\\
   \multicolumn{2}{L}{\textbf{Deterministic Parameters}}\\
   N & \text{Total number of subjects}, &
   C & \text{Available beds in hospital facilities}.\\
    \multicolumn{2}{L}{\textbf{Probabilistic Parameters}}\\
   \omega &\text{Prob. to contract the infection in one meeting}, &
   \beta &\text{Prob. for an infectious asympt. sub. to recover},\\
   \delta &\text{Prob. for an asympt. sub. to devel symptoms}, &
   \mu &\text{Prob. for a symptomatic sub. to recover},\\
   \alpha &\text{Prob. for a symptomatic sub. to die}, &
   \sigma &\text{Prob. for an hospitalised sub. to die},\\
   \xi &\text{Prob. for an hospitalised sub. to recover}, &
   \gamma &\text{Prob. for a tested infectious sub. to be positive},\\
   \psi &\text{Prob. for a symptomatic sub. to be hospitalised}, &
   \iota & \text{Prob. that a quarantined sub. devel symptoms},\\
   \upsilon &\text{Prob. that a quarantined sub. recovers}.\\
   \multicolumn{2}{L}{\textbf{Command Variables}}\\
   M_k & \text{Num. of people met by any subject}, &
   t_k & \text{Num. of people tested}.
\end{tabular}
}}
\egroup
\caption{Summary of symbols.}
\label{fig:sym}
\end{figure}

This model is based on the following assumptions:
\begin{enumerate*}[label=\roman*)]
\item the presence of a virus can be detected either if the subject
  starts to develop symptoms of the disease or when the subject is
  tested positive;
\item if a subject is tested positive (i.e. infectious) she/he becomes
  quarantiened until recovery;
\item a quarantined subject either recovers or develops the symptoms
  and becomes infectious;
\item a recovered subject cannot be re-infected;
\item since it is not possible to distinguish a subject who is
  susceptible, asymptomatic or recovered without having developed
  symptoms, the states $\mathcal{S}$, $\mathcal{A}$, $\mathcal{R}a$
  are not observable, while all the other states $\mathcal{Q}$,
  $\mathcal{I}$, $\mathcal{O}$, $\mathcal{D}$, $\mathcal{R}$ are
  observable;
\item the hospitals have a maximum capacity of $C \le N$.
\end{enumerate*}

The elements of the vector
$ \vec{V}_k = \left[S_k, A_k, I_k, R_k, O_k, D_k, Q_k, Ra_k \right]$
are subject to the following constraints:
$   S_k + A_k + I_k + R_k + Ra_k+ Q_k + O_k + D_k = N,$ 
$    O_k \leq C$.
%
We denote with
$ \vec{\Delta v} = \vec{V_{k+1}}-\vec{V_{k}}\! =\!  [ \Delta_S,
\Delta_A, \Delta_I,\Delta_R,\Delta_O,\Delta_D,\Delta_Q,\Delta_{Ra}]^T$
the change of the state vector from $\vec{V_{k}}$ to $\vec{V_{k+1}}$,
such that the input/output flow from each state of
Figure~\ref{fig:model} is respected (i.e. %
$\Delta_S = - \Delta_1$,
$\Delta_A = \Delta_1 -\Delta_2 -\Delta_3 - \Delta_9$,
$\Delta_I = \Delta_{10}+\Delta_2 -\Delta_4 -\Delta_5 -\Delta_6$,
$\Delta_R = \Delta_{4} +\Delta_8 +\Delta_{11}$,
$\Delta_O = \Delta_5 -\Delta_7 - \Delta_8$,
$\Delta_D = \Delta_6 + \Delta_7$,
$\Delta_Q = \Delta_9 - \Delta_{10} - \Delta_{11}$,
$\Delta_{Ra} = \Delta_3$). Hereafter, we will refer to these equations
with name \emph{balance equations}.\label{balanceequations}
To ensure that different subjects in the different states of this
model are non-negative we also enforce the following constraints:
$ \Delta_1 \le S_k$, $ \Delta_2 + \Delta_3 + \Delta_9 \le A_k$,
$ \Delta_{10} + \Delta_{11} \leq Q_k$,
$\Delta_4 + \Delta_5 + \Delta_6 \le I_k$, and
$\Delta_7 + \Delta_8 \le O_k$.

We denote by $l(\cdot)$ an assignment of variables:
$\Delta_i = \delta_i$, for $i = 1,\,\ldots,11$.  For an assignment of
variable $l(\cdot)$ we use $l(\cdot) \models \varphi$ to mean that the
assignment $l(\cdot)$ satisfies formula $\varphi$. For instance,
$l\left(\Delta_4=\delta_4,\,\Delta_{8}=\delta_{8},\,\Delta_{11}=\delta_{11}\right)
\models \Delta_R = \Delta_{4} +\Delta_8 +\Delta_{11}$ means that
the assignment $\delta_4,\,\delta_8,\,\delta_{22}$ to the variables
$\Delta_4$, $\Delta_8$ and $\Delta_{11}$ satisfies balance equation
$\Delta_R = \Delta_{4} +\Delta_8 +\Delta_{11}$. We also
introduce the following notations:
\begin{itemize}
\item $l_1$ is an assignment linking the variable $\Delta_1$ defined
  via $B_1$ as: $l_1$: $(\delta_1 = -\Delta_S)$;
\item $l_2$ is a function linking the variables $\Delta_2$, $\Delta_3$
  and $\Delta_9$ (with the variable $\Delta_9$ obtained via equation
  $B_2$, and the variable $\Delta_3$ obtained via equation $B_8$),
  defined as: $l_2$:
  $(\Delta_2=\delta_2, \Delta_3 = \Delta_{R_a},\,\,
  \Delta_9=\!-\!\Delta_S\!-\!\Delta_A \!-\!\Delta_{R_a} \!-\!
  \delta_2)$;

\item $l_3$ is a function linking $\Delta_4,\Delta_5,\Delta_6$ and
  given by: $l_3$:
  $(\Delta_4=\delta_4,\Delta_5=\delta_5, \Delta_6=\delta_6)$;

\item $l_4$ is an assignment linking the remaining variables defined as:\\
  $l_4$:
  $(\Delta_7 = \Delta_D - \delta_6,\,\, \Delta_8 = \delta_5 +\delta_6
  - \Delta_D - \Delta_O,\,\,
  \Delta_{10}=\Delta_I-\delta_2+\delta_4+\delta_5+\delta_6,
  \Delta_{11}=\Delta_R + \Delta_D + \Delta_O - \delta_4 -\delta_5 -
  \delta_6 )$;
\item $l_5$, finally, assigns $(\Delta_2=\delta_2, \Delta_3=\delta_3,
      \Delta_9=\delta_9)$.
\end{itemize}
Finally, we also consider the following terms:
$C_{\beta, \delta} = (1 -\beta - \delta) \ge 0$,
$C_{\mu, \psi, \alpha} = (1 -\mu - \psi -\alpha) \ge 0$,
$C_{\sigma,\xi} = (1-\sigma-\xi) \ge 0$,
$C_{\iota,\upsilon} = (1-\iota-\upsilon) \ge 0$.

The probability associated with a transition from state vector
$\vec{V}_k$ to state vector $\vec{V}_{k+1}$, such that exactly $M_k$
encounters between susceptible subjects can happen and exactly $t_k$ tests
are performed, denoted with $\Prc{\vec{V}_{k+1}}{\vec{V}_k}$ can be
computed as follows:
\begin{align}
  \Prc{\vec{V}_{k+1}}{\vec{V}_k} & = \Prc{l_1}{\vec{V}_k} \cdot \displaystyle\sum_{\delta_2=0}^{-\Delta_S-\Delta_A-\Delta_{R_a}}
    \Prc{l_2}{\vec{V}_k \wedge l_1} \nonumber\\
   & \quad \quad \cdot \displaystyle\sum_{\delta_4=0}^{\delta_2 - \Delta_I}
    \displaystyle\sum_{\delta_5=0}^{\delta_2 - \Delta_I-\delta_4} \sum_{\delta_6=0}^{\delta_2 - \Delta_I-\delta_4-\delta_5}
    \!\!\!\!\!\!  \Prc{l_3}{\vec{V}_k \wedge l_1 \wedge l_2}  \label{eq:transProbFinalExt}\\
   &  \quad \quad \quad \cdot \Prc{l_4}{\vec{V}_k \wedge l_1 \wedge l_2 \wedge l_3}\!\! \nonumber
\end{align}
where:
\begin{itemize}[topsep=0pt]
\item $\Prc{l_1}{\vec{V}_k}$ is the probability that exactly $\delta_1$
susceptible become asymptomatic, and is computed as
$\Prc{l_1}{\vec{V}_k} = \mathcal{B}(S_k,
\Prc{g_k}{\vec{V}_k}_{M_k})_{-\Delta_{S}}$ where, and
$\Prc{g_k}{\vec{V}_k}_{M_k} = 1 - \left( \!1\! - \!  \frac{\omega
    A_k}{N\!-\!D_k\!-\!I_k\!-\!O_k\!-\!Q_k} \right)^{M_k}$.

\item $\Prc{l_2}{\vec{V}_k \wedge l_1}$ is defined as
$\displaystyle \sum_{H = \delta_9}^{t_k} \Pr{s_H} \sum_{F =
  0}^{\delta_9} \rho(\delta_2, \delta_3\!+\!F, \vec{V}_k)
\binom{A_k}{H}^{-1} K$ where $t_k$ are the tests performed in the
transition from $ \vec{V}_k$ to $\vec{V}_{k+1}$,
$ K = \binom{\delta_3+F}{F} \binom{A_k - (\delta_2 + \delta_3 +
  F)}{\delta_9 - F} \binom{\delta_2}{H - \delta_9}$,
$\rho(\delta_2, \delta_3, \vec{V}_k) = \frac{A_k!\, \beta^{\delta_3}\,
  \delta^{\delta_2}\, C_{\beta,
    \delta}^{A_k-\delta_2-\delta_3}}{\delta_2! \delta_3!
  (A_k-\delta_2-\delta_3)!}$, and
$\Pr{s_{H}} = \binom{N_k}{t_k}^{-1} \sum_{p = 0}^{t_k} \binom{S_k +
  R_{a_k}}{t_k - p} \binom{A_k}{p} \mathcal{B}(p,\gamma)_H$.
\item $\Prc{l_3}{\vec{V}_k \wedge l_1 \wedge l_2}$ is defined as
\[
\begin{aligned}
\Prc{l_3}{\vec{V}_k \wedge l_1 \wedge l_2} = & \begin{cases}
0 & \!\!\!\text{if }\delta_4+\delta_5+\delta_6 > I_k \vee \delta_5 > C - O_k \\
M(\delta_4, \delta_5, \delta_6) & \!\!\!\text{if } \delta_5 < C - O_k \wedge \delta_4+\delta_5+\delta_6 \leq I_k \\
M'(\delta_4, \delta_6)  & \!\!\!\text{if } \delta_5 = C - O_k \wedge \delta_4+\delta_5+\delta_6 \leq I_k
\end{cases}
\end{aligned}
\]
where
$M(\delta_4, \delta_5, \delta_6) =
\mathbb{M}_{I_k,\delta_4,\delta_5,\delta_6} \mu^{\delta_4}\,
\psi^{\delta_5}\, \alpha^{\delta_6}\, C_{\mu, \psi, \alpha}^{I_k -
  \delta_4 - \delta_5 - \delta_6} $ and
$M'(\delta_4, \delta_6) = \sum_{h = 0}^{I_k - \delta_4 - \delta_6
  -(C - O_k)} M(\delta_4, (C - O_k)+h, \delta_6)$;
\item $\Prc{l_4}{\vec{V}_k \wedge l_1 \wedge l_2 \wedge l_3}$ is
  defined as
  $\zeta(\Delta_D\! -\! \delta_6, \delta_5\! +\!\delta_6\! -\!
  \Delta_D \!-\! \Delta_O, \vec{V}_k) \cdot \chi(\Delta_I \!-\!
  \delta_2\!+\!\delta_4\!+\!\delta_5\!+\!\delta_6, \Delta_R \!+\!
  \Delta_D \!+\! \Delta_O \!-\! \delta_4 \!-\!\delta_5 \!-\! \delta_6,
  \vec{V}_k)$ where
  $\zeta(\delta_7,\delta_8, \vec{V}_k) =
  \mathbb{M}_{O_k,\delta_7,\delta_8} \sigma^{\delta_7}\,
  \xi^{\delta_8}\, C_{\sigma,\xi}$, and
  $\chi(\delta_{10},\delta_{11}, \vec{V}_k) =
  \mathbb{M}_{Q_k,\delta_{10},\delta_{11}} \iota^{\delta_{10}}\,
  \upsilon^{\delta_{11}}\,
  C_{\iota,\,\upsilon}^{Q_k\!-\!\delta_{10}\!-\!\delta_{11}}$.
\end{itemize}

\noindent All the mathematical details and proofs to show the
correctness of the above formulation for computing the probability of
a transition from $\vec{V}_k$ to $\vec{V}_{k+1}$ subject to having
exactly $M_k$ meetings and performing exactly $t_k$ tests for the
model depicted in Figure~\ref{fig:model} are out of the scope of this
paper and can be found in~\cite{DBLP:journals/corr/abs-2204-11317}.

Given the above definitions, we can compute the transitions and the
associated probabilities from a state vector $\vec{V_k}$ given exactly
$M_k$ encounters, and exactly $t_k$ tests by enumerating all possible
configurations $\vec{V_{k+1}}$ that are compatible with the
\emph{balance equations} at page
\pageref{balanceequations}. Algorithm~\ref{alg:transitionalgorithm0}
in Appendix \ref{sec:algos} shows how to perform this
enumeration.

The transitions from a state vector $\vec{V_k}$ subjected to
encounters from a set of minimum encounters $M_{min}$ to a maximum of
$M_{max}$ encounters, and tests from a minimum of $T_{min}$ to a
maximum of $T_{max}$ can be computed by enumerating all possible
$(m,t)$ such that $m \in [M_{min},M_{max}]$ and
$t \in [T_{min},T_{max}]$ using previous algorithm (see
Algorithm~\ref{alg:transitionalgorithm1} in Appendix \ref{sec:algos}
for details).
These algorithms
are the building blocks for computing the MDP for the full stochastic
model for the spread of a \sarscov{}-like infection for a population
of size $N$ such that each subject evolves as illustrated in
Figure~\ref{fig:model}.
The set of states of the MDP are all those vector states
$\vec{V}_k = \left[S_k, A_k, I_k, R_k, O_k, D_k, Q_k, Ra_k \right] $
such that they satisfy the constraint
$S_k + A_k + I_k + R_k + O_k + D_k + Q_k + Ra_k = N$ for a population
of $N$ subjects%
\footnote{As shown in \cite{DBLP:journals/corr/abs-2204-11317}, this model is such that, for a
population of $N$ subjects, assuming there are $n$ possible
configurations (in our case $n=8$), the maximum number of possible
states that can be generated is $\binom{N+n-1}{N}$ that corresponds to
the Bose-Einstein statistics.}.
The set of actions can be computed as
$A = \{\langle M_k,t_k \rangle \;| \; M_k \in [M_{min},M_{max}], t_k
\in [t_{min},t_{max}]\}$, for the possibility to meet subjects from
$M_{min}$ to $M_{max}$, and to perform tests from $t_{min}$ to
$t_{max}$. The transition probability function
$T = \{ \langle \vec{V}_k, \langle M_k, t_k \rangle, \vec{V}_{k+1},
\Prc{\vec{V}_{k+1}}{\vec{V}_k} \rangle \; | \langle M_k, t_k \rangle
\in A, \vec{V}_{k} \in S, \langle \vec{V}_k, \langle M_k, t_k \rangle,
\vec{V}_{k+1}, \Prc{\vec{V}_{k+1}}{\vec{V}_k} \rangle \in
\textsc{Transitions}(\vec{V}_k, M_k, t_k)\}$. Finally,
$S_I \subseteq S$ is the set of initial states. In this model, we do
not consider any reward function.

This framework allows for the application of several policies to
control the (PO)MDP. To this extent, we see a policy as a function
$\mathcal{P}(\vec{V}_k) \rightarrow 2^A$ that associates with a state
$\vec{V}_k$ a pair $\langle M_k, t_k \rangle$ such that
$M_k \in [M_{min},M_{max}]$ and $t_k \in [t_{min},t_{max}]$. This can
be achieved by restricting the transition probability function to
follow policy $\mathcal{P}$ as follows:
$T = \{ \langle \vec{V}_k, \langle M_k, t_k \rangle, \vec{V}_{k+1},
\Prc{\vec{V}_{k+1}}{\vec{V}_k} \rangle \; | \langle M_k, t_k \rangle
\in \mathcal{P}(\vec{V}_k) \subseteq A, \vec{V}_{k} \in S, \langle
\vec{V}_k, \langle M_k, t_k \rangle, \vec{V}_{k+1},
\Prc{\vec{V}_{k+1}}{\vec{V}_k} \rangle \in
\textsc{Transitions}(\vec{V}_k, M_k,
t_k)\}$. Algorithms~\ref{alg:transitionalgorithm0} and
\ref{alg:transitionalgorithm1} can be easily adapted to restrict the
actions to obey a given policy $\mathcal{P}(\vec{V}_k)$. In
particulara, it is sufficient in Algorithm
\ref{alg:transitionalgorithm1} to replace the two nested for loops
with a single loop over elements of a set of pairs
$\langle m,t \rangle \in \mathcal{P}(\vec{V}_k)$.


%% file: experimental.tex
Once the (PO)MDP model has been built, one can convert it into the input
format of a stochastic model checker like e.g. PRISM~\cite{kwiatkowska2011prism} or STORM~\cite{DBLP:journals/corr/abs-2002-07080}, and use this model to
verify PCTL~\cite{HJ94} properties, for synthesizing policies satisfying a
given PCTL property, to evaluate formally the effects of a policy, and to compute steady state probabilities.

To this extent, we have developed a C++ proof of concept open-source
tool, name
\texttt{covid\_tool}\footnote{\url{https://bitbucket.org/luigipalopoli/covd_tool}}. As
an input, this tool receives (encoded in a json file) the population
size $N$, all the probability parameters described in
Figure~\ref{fig:sym}, the $M_{min}$, $M_{max}$, $t_{min}$, $t_{max}$
and the hospital capacity $C$. The tool also accepts in input a set of
possible initial states where for each initial state is fully
specified by the respective $\vec{V}$.
Moreover, to evaluate the possible effect of manually specified
control policies, we integrated in the tools four policies
$\mathcal{P}_{-1}$, $\mathcal{P}_0$, $\mathcal{P}_1$, and
$\mathcal{P}_2$.
$\mathcal{P}_{-1}$ corresponds to not applying any policy, and this
results in generating all possible pairs in $A$ (see e.g.,
Algorithm~\ref{alg:transitionalgorithm1}).  $\mathcal{P}_{0}$ is a
constant policy that regardless of the state $\vec{V}_k$ returns a
singleton element $\langle M,t \rangle$ (i.e.,
$\forall \vec{V}_k \in S. \mathcal{P}_0(\vec{V}_k) = \{\langle M, t
\rangle\}$).  Policies $\mathcal{P}_1$ and $\mathcal{P}_2$ have the
following form
$\mathcal{P}(\vec{V}_k) = \{\langle M_k, t\rangle | M_k =
\mathcal{F}(\vec{V}_k), t \in [t_{min},t_{max}]\}$ where
\begin{equation}
\mathcal{F}(\vec{V}_k)\! =\! \begin{cases}
  M_{max}   &  \text{if } f(\vec{V}_k) \leq T_{\downarrow}\\
  M_{min}   &  \text{if } f(\vec{V}_k) \geq T_{\uparrow}\\
  M_{max}\! +\! \left(M_{min}\!-\!M_{max}\right) \frac{f(\vec{V}_k)-T_{\downarrow}}{T_{\uparrow}-T_{\downarrow}} & \text{otherwise}.
  \end{cases}\!\!
\end{equation}
$\mathcal{P}_1$ uses $f(\vec{V}_k)= I_k + O_k/(N-D_k)$ while
$\mathcal{P}_2$ uses $f(\vec{V}_k) = A_k/(N-D_k)$.  In $\mathcal{P}_1$
when the percentage of the number of symptomatic and hospitalized
patients over the living population is below the threshold
$T_{\downarrow}$, we impose no restrictions for social life. If this
number is above a threshold $T_{\uparrow}$, we adopt the maximum
restriction (the minimum value $M_{min}$ for $M$). Otherwise we adopt
a linear interpolation between the minimum and the maximum values of
$M$. $\mathcal{P}_2$ is similar to $\mathcal{P}_1$, but here we
consider the ratio between asymptomatic infected subjects and the
living population.
The tool and all the material to reproduce the experiments
reported hereafter are available at
\url{https://bitbucket.org/luigipalopoli/covd_tool}.

This tool uses Algorithms~\ref{alg:transitionalgorithm0} and
\ref{alg:transitionalgorithm1} to build the MDP for the model of
Figure~\ref{fig:model}, and the respective adaptation of such
algorithms to generate the DTMCs resulting from the application of a
given pre-defined policy $\mathcal{P}$. Among the different
possibilities this tool provides, we highlight here the ability to
generate a (PO)MDP symbolic model as accepted by PRISM and STORM. The
symbolic model encodes the state vector $\vec{V}$ with 8 integer
variables (\texttt{S}, \texttt{A}, \texttt{I}, \texttt{R}, \texttt{O},
\texttt{D}, \texttt{Q}, \texttt{Ra}) with values ranging from 0 to
$N$. We encode actions with a label \texttt{action\_m\_t} for meeting
exactly $m$ subjects and performing exactly $t$ tests. Then using
Algorithm~\ref{alg:transitionalgorithm0} to compute the possible next
states and respective probabilities.
The tool is also able to generate a symbolic Partially Observable
Markov Decision Problems in PRISM format by specifying that the
\texttt{S}, \texttt{A}, \texttt{Ra} are not observable.
Moreover, the tool can handle two cases, the full model with all the
eight states as per Figure~\ref{fig:model}, and a simplified model
that does not considers quarantined and the possibility to recover
from asymptomatic state (that corresponds to a SAIROD
model). We introduced this possibility for two main reasons. First, the SAIROD model has been already studied, and it is easier to retrieve the parameters governing the behavior~\cite{giordano:2020}. Second, as we will see later on, the full model is subject to scalability issues much quickly than the simple SAIROD one. Listing~\ref{simpleprism} is an excerpt of a simple PRISM
model corresponding to a population composed of 10 subjects.

\begin{lstlisting}[language=prism,escapeinside={(*}{*)},caption=Excerpt of a PRISM file generated for a SAIROD model with a population of 10 individuals. The entire file is about 103Mb on disk.,float=t!,label=simpleprism]
  mdp // The kind of model
  module covid_mdp // The main module
    // The variables
    nS : [0..10]; nA : [0..10]; nI : [0..10]; nR : [0..10]; nO : [0..10]; nD : [0..10];
    ... // omitted for lack of space
    [act_M_is_5]  ((nS = 0) & (nA = 0) & (nI = 0) & (nR = 0) & (nO = 1) & (nD = 9)) ->
                  0.5 : (nS' = 0) & (nA' = 0) & (nI' = 0) & (nR' = 1) & (nO' = 0) & (nD' = 9)
                + 0.2 : (nS' = 0) & (nA' = 0) & (nI' = 0) & (nR' = 0) & (nO' = 1) & (nD' = 9)
                + 0.3 : (nS' = 0) & (nA' = 0) & (nI' = 0) & (nR' = 0) & (nO' = 0) & (nD' = 10);
    ... // omitted for lack of space
  endmodule
  init // The set of initial states
     (  (nS = 7) & (nA = 3) & (nI = 0) & (nR = 0) & (nO = 0) & (nD = 0)
      | (nS = 9) & (nA = 1) & (nI = 0) & (nR = 0) & (nO = 0) & (nD = 0) )
  endinit
  \end{lstlisting}

\begin{table}[b!]
\centering  \scalebox{0.85}{
\begin{tabular}{ |m{0.5cm} |m{0.5cm}| m{1cm}| m{1cm} | m{2.5cm}| m{3.8cm} |m{3.8cm} | }
  \hline
    $N$  & $C$ & $P_{\textit{min}}$ & $P_{\textit{max}}$ & Creation of $T$ (s) & STORM model creation (s) & STORM model checking (s) \\
  \hline
   5 & 1 & 0.250 & 0.312 & 0.088& 0.297 & 0.005 \\

   10 & 1 & 0.260 & 0.346 & 25.161 &  58.705 & 56.955 \\

   15 & 1 & 0.264 & 0.382 & 1118.465 & 3205.290 & 33.257 \\

   20 & 1 &  & memout &  &  &   \\

   10 & 2 & 0.005 & 0.019 & 36.080 &  103.292 & 2.191 \\

   15 & 3 &  &  & memout &  &   \\

  \hline
\end{tabular}
}
\caption{\label{tab:results_mdp_esv_cluster} Experiments using the extended model, no policy ($\mathcal{P}_{-1}$).}
\end{table}

The generation of models in PRISM language is subject to severe
efficiency and espressivity problems.  First, the PRISM language can
represent symbolically the transitions once the
$\Prc{\vec{V}_{k+1}}{\vec{V}_k}$ are pre-computed for each transition,
but there is no efficient way to encode
$\Prc{\vec{V}_{k+1}}{\vec{V}_k}$ in the language due to the limited
espressivity of the PRISM language (the involved math is not supported
by the language). A possibility that we considered was to build a
defined symbols to represent the $\Prc{\vec{V}_{k+1}}{\vec{V}_k}$ for
all possible values of the $\vec{V}_k$, $M_k$, and $t_k$, however the
resulting file will be huge (larger than a gigabyte) even for a very
small population ( $<10$ subjects). Thus, we ended up pre-computing such probabilities, and associating the resulting value with each transition. The size of the file is problematic for two reasons: first problem is a storage problem. Second problem, assuming the huge file has been generated successfully, the PRISM and STORM model checkers require a large amount of memory and huge timing (days) to parse the file before even starting verification on modern high-performance computers equipped
with large memory (Terabytes). Here we remark that STORM is slightly more efficient than PRISM in handling large input files. This might be due to a number of reasons, notably the fact that STORM uses an efficient C++ parser, while PRISM uses Java. To overcome these limitations, we considered two directions. First, we considered the possibility to generate the explicit transition matrix files in the different formats accepted by the STORM and PRISM model checkers (the generated files with this flow are smaller than the symbolic approach, but still large and requiring large computation and memory to handle them). Second, we followed the direction to have a more strict integration with the model checker by
linking the model checker in memory (thus avoiding intermediate file
generation). In particular, we did a tight integration with the STORM
model checker by directly building in memory the data structures to
enable model checking. We have chosen STORM for two main reasons. Fist, it is written in C++ while PRISM is written in Java. Second, STORM provides a clear API to facilitate the integration in other tools. Currently, we are only building the sparse matrix representation~\cite{DBLP:journals/corr/abs-2002-07080}, and
thus we are limited to the verification capabilities  by STORM with this model representation. 
(See \url{https://www.stormchecker.org/} for further details.)

All the experiments have been executed on a cluster equipped with 112
Intel(R) Xeon(R) CPU cores \@ 2.20GHz and 256Gb of RAM. We considered
a memory limit of 256GB and a CPU time limit of 5400
seconds.


\begin{table}[t!]
\centering    \scalebox{0.85}{
\begin{tabular}{ | m{0.4cm} | m{0.9cm}| m{0.5cm} | m{0.5cm} | m{1cm} | m{2.5cm}| m{3.8cm} |m{3.8cm} | }
  \hline
  & Policy & $N$  & $C$ & $P$ & Creation of $T$ (s) & STORM model creation (s) &  STORM model checking (s) \\
  \hline

   \parbox[t]{2mm}{\multirow{6}{*}{\rotatebox[origin=c]{90}{ConstLow}}} & $\mathcal{P}_{0}$ & 5 & 1 & 0.280 & 0.088 & 0.178 & 0.001 \\

     & $\mathcal{P}_{0}$ & 10 & 1 & 0.300 & 25.111 & 33.044 & 0.170 \\

     & $\mathcal{P}_{0}$ & 15 & 1 & 0.308 & 1131.249 & 988.710 & 5.031 \\

     & $\mathcal{P}_{0}$ & 10 & 2 & 0.011 & 36.010 & 49.453 & 0.301 \\

     & $\mathcal{P}_{0}$ & 15 & 3 & 0.0006 & 2371.702 & 2179.716 & 16.505 \\

     & $\mathcal{P}_{0}$ & 20 & 4 &  & memout & & \\

    \hline

    \parbox[t]{2mm}{\multirow{6}{*}{\rotatebox[origin=c]{90}{ConstHigh}}} & $\mathcal{P}_{0}$ & 5 & 1 & 0.280 & 0.088 & 0.175 & 0.001 \\

     & $\mathcal{P}_{0}$ & 10 & 1 & 0.300 & 25.032 & 33.642 & 0.182  \\

     & $\mathcal{P}_{0}$ & 15 & 1 & 0.308 & 1130.826 & 956.313 & 5.050  \\

     & $\mathcal{P}_{0}$ & 20 & 1 & & timeout & &   \\

     & $\mathcal{P}_{0}$ & 10 & 2 & 0.011 & 35.977 & 48.314 & 0.304 \\

     & $\mathcal{P}_{0}$ & 15 & 3 & 0.308 & 1130.826 & 956.313 & 5.051 \\

     & $\mathcal{P}_{0}$ & 20 & 4 & & timeout &  &  \\

    \hline

    \parbox[t]{2mm}{\multirow{5}{*}{\rotatebox[origin=c]{90}{Adapt1}}} & $\mathcal{P}_{1}$ & 5 & 1 & 0.280 & 0.088 & 0.179 & 0.001 \\

     & $\mathcal{P}_{1}$ & 10 & 1 & 0.300 & 25.279 & 34.985 & 0.181 \\

     & $\mathcal{P}_{1}$ & 15 & 1 & 0.308 & 1132.725 & 957.143 & 5.005 \\

     & $\mathcal{P}_{1}$ & 20 & 1 & & memout & &  \\


     & $\mathcal{P}_{1}$ & 10 & 2 & 0.011 & 35.885 & 49.046 & 0.301 \\

     & $\mathcal{P}_{1}$ & 15 & 3 & & & timeout &  \\

    \hline

    \parbox[t]{2mm}{\multirow{6}{*}{\rotatebox[origin=c]{90}{Adapt1}}} & $\mathcal{P}_{2}$ & 5 & 1  & 0.280 & 0.088 & 0.187 & 0.001 \\

     & $\mathcal{P}_{2}$ & 10 & 1 & 0.300 & 25.062 & 32.684 & 0.184  \\

     & $\mathcal{P}_{2}$ & 15 & 1 & 0.308 & 1124.658 & 968.644 & 5.735  \\

     & $\mathcal{P}_{2}$ & 20 & 1 & & memout & &  \\


     & $\mathcal{P}_{2}$ & 10 & 2  & 0.011 & 35.931 & 48.781 & 0.306 \\

     & $\mathcal{P}_{2}$ & 15 & 3  & 0.0006 & 2528.367 & 2473.946 & 15.958 \\

     & $\mathcal{P}_{2}$ & 20 & 4  & & timeout & &  \\

    \hline

    \parbox[t]{2mm}{\multirow{6}{*}{\rotatebox[origin=c]{90}{Adapt2}}} & $\mathcal{P}_{1}$ & 5 & 1  & 0.280 & 0.087 & 0.179 & 0.001 \\

     & $\mathcal{P}_{1}$ & 10 & 1  & 0.300 & 25.048 & 32.583 & 0.169 \\

     & $\mathcal{P}_{1}$ & 15 & 1  & 0.308 & 1134.407 & 974.875 & 5.013  \\

     & $\mathcal{P}_{1}$ & 20 & 1 &  & memout &  &  \\


     & $\mathcal{P}_{1}$ & 10 & 2  & 0.011 & 35.885 & 48.797 & 0.316 \\

     & $\mathcal{P}_{1}$ & 15 & 3 & & & timeout &  \\

    \hline

    \parbox[t]{2mm}{\multirow{6}{*}{\rotatebox[origin=c]{90}{Adapt2}}} & $\mathcal{P}_{2}$ & 5 & 1  & 0.280 & 0.088 & 0.183 & 0.001 \\

     & $\mathcal{P}_{2}$ & 10 & 1  & 0.300 & 25.080 & 32.528 & 0.181 \\

     & $\mathcal{P}_{2}$ & 15 & 1  & 0.308 & 1124.280 & 949.528  & 5.051 \\

     & $\mathcal{P}_{2}$ & 20 & 1 & & memout & &  \\


     & $\mathcal{P}_{2}$ & 10 & 2 & 0.280 & 0.088 & 0.347 & 0.001 \\

     & $\mathcal{P}_{2}$ & 15 & 3 &  & & timeout &   \\

  \hline
\end{tabular}
}
\caption{\label{tab:results_dtmc_esv_cluster} Experiments using the extended model, policies $\mathcal{P}_{0}$, $\mathcal{P}_{1}$ and $\mathcal{P}_{2}$.}
\end{table}


We conducted experiments with varying population sizes $N$, hospital capacities $C$ and policies used. The values of $M_{\textit{min}}$ and $M_{\textit{max}}$ used were 1 and 5, respectively. The values of $t_{\textit{min}}$ and $t_{\textit{max}}$ used were 1 and 3, respectively. These values are reasonable for the population sizes we managed to consider (see the results later in this section).
All the values of the parameters (e.g., transition probabilities) were based on discussions with experts. The precise values used in all these experiments can be found in the aforementioned bitbucket Git repository of the tool.
We performed the experiments on two versions of the model:
\begin{enumerate*}[label=\roman*)]
\item the ``\emph{full}'' model described in previous section (the results are reported in Tables ~\ref{tab:results_mdp_esv_cluster} and ~\ref{tab:results_dtmc_esv_cluster}), and
\item the ``\emph{reduced}'' model, which does not consider the possibility of entering in quarantine ($\mathcal{Q}$) and the possibility for an asymptomatic to recover ($\mathcal{R}a$) (the results are reported in Tables~~\ref{tab:results_mdp_ssv_cluster} and ~\ref{tab:results_dtmc_ssv_cluster} in the appendix).
\end{enumerate*}
Moreover, we also considered the effects of the different considered policies.
Tables ~\ref{tab:results_mdp_esv_cluster} and ~\ref{tab:results_mdp_ssv_cluster} report the results for $\mathcal{P}_{-1}$,
while Tables ~\ref{tab:results_dtmc_esv_cluster} and ~\ref{tab:results_dtmc_ssv_cluster} show cases for policies $\mathcal{P}_{0}$, $\mathcal{P}_{1}$ and $\mathcal{P}_{2}$.
In the experiments with policies $\mathcal{P}_{0}$, $\mathcal{P}_{1}$, $\mathcal{P}_{2}$ we considered the verification of the PCTL formula $\PProb_{ = ?} [ \LEventually O = C ]$ (i.e. the probability of eventually reaching a state in which the hospital is saturated), while in the experiments with $\mathcal{P}_{0}$, we find minimum and maximum probabilities ($\PProb_{ min = ?} [ \LEventually O = C ]$ and $\PProb_{ max = ?} [ \LEventually O = C ]$, respectively). These properties were chosen given the interest of experts and decision makers of knowing the probabilities to saturate hospitals in different conditions.
Constant policies ConstHigh and ConstLow use the $M$ values of 1 and 5, respectively.
With Adapt1 we denote the adaptive policy with $T_{\downarrow} = 0.1$ and $T_{\uparrow} = 0.5$. Adapt2 denotes the adaptive policy with $T_{\downarrow} = 0.05$ and $T_{\uparrow} = 0.15$.

For each experiment we report:
\begin{enumerate*}[label=\roman*)]
\item the time in seconds required to build the complete transition matrix with the approach described in the previous section; 
\item the time in seconds to fill and build the model in memory within the STORM model checker;
\item the time in seconds required by STORM to model check the given property on the previously built model;
\item the computed probabilities for the considered properties.
\end{enumerate*}

The results in the tables clearly show that the time is mostly divided between computing the transition probability matrix and creating the STORM model (with checking the property taking relatively little time). This is due to the large number of states and transitions even for the small population sizes considered.
With the hardware at our disposal, we mostly manage to deal with population sizes up to 25. All the experiments ran out of memory with larger values of $N$ while computing the transition probability matrix. We spent a significant engineering effort to limit this explosion trying to find efficient methods to represent states and transitions, as well as memorizing the result of the computation of the transition probabilities. However, despite this engineering effort, the large state space required reached easily the limits of the hardware at our disposal. We remark that, it is in principle possible to address scalability to large population size by considering a unit of population in the model as the representative of a (larger) number of people with a numerically quantifiable error (a similar approach has been discussed in \cite{nasir2020epidemics}, and is left as future work).
%
The results also shows that, as expected, the probability of hospitals being saturated increases with increasing population sizes and decreases with greater hospital capacities.

We remark that, despite the limited scalability issues we encountered, this work constitute a basis to challenge stochastic model checkers along different directions (e.g., expressivity to allow for a concise representation of cases like this one, and efficiency to allow handle more realistic size scenarios). Moreover, it opens to the possibility to leverage the feature provided by stochastic model checkers to compute policies to achieve given properties of interest for a decision maker (although we have not yet experimented with this feature, and we will leave as future work).



%% file: rw.tex
There have been several works that addressed the problem of
mathematically modeling the spread of
diseases~\cite{bernoulli:1760,Kermack:1927,allen:2008,Brauer:2019,giordano:2020,ghezzi:1997,giordano:2016,Khanafer:2016,Yousefpour:2020}\footnote{We
  refer the reader to \cite{DBLP:journals/corr/abs-2204-11317} for a more detailed discussion
  of the literature on modeling and analyzing the spread of diseases
  with analytical models.}. These works consider models where the
population has been break down into several compartments like e.g. the
Susceptible-Infected-Recovered (SIR) model which is a simplified
version w.r.t. the one adopted in this paper.
Some of the recent works (like e.g.~\cite{allen:2008,Brauer:2019})
focused on analyzing strategies to keep in check the evolution of the
epidemic leveraging the control variables with the goal to construct
interesting control theoretical results. For instance,
in~\cite{giordano:2020} has been presented a model with many
compartments. In~\cite{Khanafer:2016} has been analyzed the problem of
stability, while policies for \covid{} based on Optimal Control are
discussed in~\cite{Yousefpour:2020}.
All of these models are deterministic and aim at capturing the disease
dynamics. Stochastic models, differently from deterministic ones,
allows to derive richer set of informations.
For instance, stochastic models
\begin{enumerate*}[label=\roman*)]
\item may converge to a disease-free state even if the corresponding
  deterministic models converge to an endemic
  equilibrium~\cite{anderson:1992};
\item may allow for computing the probability of an outbreak, the
  distribution of the final size of a population or the expected
  duration of an epidemic (see e.g.~\cite{Brauer:2019,Sattenspiel:1990});
\item may allow to quantify the probability of transition between
  different state of \covid{}-affected patients based on the age class
  (see e.g.~\cite{ZARDINI2021100530});
\item allow to evaluate the effects of lock-down policies (see
  e.g.~\cite{Flavia20}).
\end{enumerate*}
An important class of models amenable to analytical analysis are
Markov Processes~\cite{cassandras2009introduction,allen:2008}. When
we observe the system in discrete--time, Markov Models are called
discrete-time Markov chains (DTMC). When command variables become part
of the model, Markov Models are called Markov Decision Processes
(MDP), and where not all states are directly observable (e.g.,
asymptomatic persons), we have a Partially Observable MDPs (POMDP).
These models, contrary to other stochastic models such as Stochastic
Differential Equations (SDE), adopt a numerable state space composed
of discrete variables.

In the literature two paradigms have been adopted to model a disease
spread as a DTMC, namely the Reed-Frost model and the Greenwood model
~\cite{gani:1971,Tuckwell:2007}. In all these models the transition
probabilities are governed by binomial random variables.  Extensions
of this model were presented by.
The use of stochastic models opens for the possibility to use
Stochastic Model Checking in order to study probabilistic temporal
properties to evaluate the effects of a strategy on a population
during the evolution of a disease~\cite{10.1371/journal.pone.0145690,C1MB05060E,Chauhan2015}.
An adapted version of the
Susceptible-Exposed-Infectious-Recovered-Delayed-Quarantined
(Susceptible/Recovered) continuous time Markov chain model has been
used in~\cite{10.1371/journal.pone.0145690} to analyze the spread of
internet worms using the PRISM model
checker~\cite{kwiatkowska2011prism}.
A stochastic model to compute with the PRISM model checker the minimum
number of influenza hemagglutinin trimmers required for fusion to be
between one and eight has been proposed in~\cite{C1MB05060E}.
The use of stochastic simulations to compute timing parameters for a
timed automaton has been studied in~\cite{Chauhan2015}.
All these stochastic models are rather simplified and abstract models
of the disease spread, and the main reason for such is
tractability. Indeed, considering large models with complex dynamics
(as shown in this paper) reach quickly to computation limits even on
recent computation infrastructures. Moreover, all these models, to
make the model tractable by model checkers, enforce that only one
subject can change her/his state across one transition or do not
consider command variables. In this work, leveraging on the stochastic
model defined in~\cite{DBLP:journals/corr/abs-2204-11317} we allow for multiple subjects to
change state simultaneously across one transition, and we allow for
command variables. In this work we show the limits of this more
realistic model and show challenges for making next generation
stochastic model checkers suitable for analyzing complex disease
stochastic models.

The problem of scalability of epidemic models has been discussed
in~\cite{nasir2020epidemics,Hak18}. The work in
\cite{nasir2020epidemics} addressed the scalability to large
population size by considering a unit of population in the model as
the representative of a number of people. \cite{Hak18} addresses the
problem by considering a graph of MDPs, each governed by the same
update rules, that interact with their neighbors following the given
graph topology. It would be interesting to see how verification
techniques could leverage these abstractions to address the
scalability issues. However, this is left to future work.


%% file: conclusions.tex

In this paper we considered the study of an epidemic model for the
evolution of diseases modeled with stochastic models in form of Markov
Processes, and we showed how to encode such complex model into
formalisms suitable for being analyzed with state-of-the-art
stochastic model checkers. We developed an open source tool that given
the parameters of the epidemic model is able to generate models in the
PRISM formalism (a widely used formalism), as well as to build
directly in memory the STORM sparse model by linking our tool with the
STORM model checker. We used the developed tool to study the model
with increasing population sizes, analyzing the models against given
temporal properties, and evaluating the effects of different control
policies w.r.t.\ some interesting temporal properties. The results
showed that the approach is feasible, but it is subject to scalability
issues even with small population sizes. Moreover, this work
highlighted several challenges for the community to address large
population sizes on one hand, and espressivity requirements on the
input languages on the other hand to simplify the specification of
such complex mathematical models.

As future work, we want to investigate the use of abstraction
techniques to improve the performance and to handle large population
sizes. Moreover, we aim also to leverage the framework to synthesize
policies with a clear guarantee on the respective effects. In terms of
modeling, the model could be further extended to consider the
vaccinated population as well as vaccinations.


%% file: appendix.tex
\section{Algorithms for computing the transition probabilities}
\label{sec:algos}

The algorithms to compute the transitions and the associated
probabilities from a state vector $\vec{V_k}$ given exactly $M_k$
encounters, and exactly $t_k$ tests.
\begin{algorithm}[tbh]
\caption{\textsc{Transitions}($\vec{V}_k, M_k, t_k$) : Transitions from state $\vec{V_k}$ subjected to $M_k$ encounters, and $t_k$ tests.}\small
\label{alg:transitionalgorithm0}
\scalebox{0.9}{
\begin{minipage}{1.2\textwidth}
\begin{algorithmic}[1]
\REQUIRE $\vec{V}_k = \left[S_k, A_k, I_k, R_k, O_k, D_k, Q_k, Ra_k \right]$, $M_k$, $t_k$
\ENSURE $Trans$
\STATE $Trans=\emptyset$
\FOR{$\delta_1 = 0$ \TO $S_k$}
\FOR{$\delta_2 = 0$ \TO $A_k$}
\FOR{$\delta_3 = 0$ \TO $A_k - \delta_2$}
\FOR{$\delta_4 = 0$ \TO $I_k$}
\FOR{$\delta_5 = 0$ \TO $I_k - \delta_4$}
\FOR{$\delta_6 = 0$ \TO $I_k - \delta_4 - \delta_5$}
\FOR{$\delta_7 = 0$ \TO $O_k$}
\FOR{$\delta_8 = 0$ \TO $O_k - \delta_7$}
\FOR{$\delta_9 = 0$ \TO $A_k - \delta_2 - \delta_3$}
\FOR{$\delta_{10} = 0$ \TO $Q_k$}
\FOR{$\delta_{11} = 0$ \TO $Q_k - \delta_{10}$}
\STATE $S_{k+1} = S_k - \delta_1$
\STATE $A_{k+1} = A_k+\delta_1-\delta_2-\delta_3-\delta_9$
\STATE $I_{k+1} = I_k+\delta_{10}+\delta_2-\delta_4-\delta_5-\delta_6$
\STATE $ R_{k+1} = R_k+\delta_4+\delta_8+\delta_{11}$
\STATE $ O_{k+1} = O_k+\delta_5-\delta_7-\delta_8$
\STATE $ D_{k+1} = D_k+\delta_6+\delta_7$
\STATE $ Q_{k+1} = Q_k+\delta_9-\delta_{10}-\delta_{11}$
\STATE $ Ra_{k+1} = Ra_k+\delta_3$
\STATE $\vec{V}_{k+1} = \left[S_{k+1}, A_{k+1}, I_{k+1}, R_{k+1}, O_{k+1}, D_{k+1}, Q_{k+1}, Ra_{k+1} \right]$
\STATE $Trans = Trans \cup \langle \vec{V}_k, \langle M_k,t_k\rangle, \vec{V}_{k+1}, \Prc{\vec{V}_{k+1}}{\vec{V}_k}\rangle$
\IENDFOR{}
\IENDFOR{}
\IENDFOR{}
\IENDFOR{}
\IENDFOR{}
\IENDFOR{}
\IENDFOR{}
\IENDFOR{}
\IENDFOR{}
\IENDFOR{}
\IENDFOR{}
\RETURN $Trans$
\end{algorithmic}
\end{minipage}
}
\end{algorithm}

The transitions from a state vector $\vec{V_k}$ subjected to
encounters from a set of minimum encounters $M_{min}$ to a maximum of
$M_{max}$ encounters, and tests from a minimum of $T_{min}$ to a
maximum of $T_{max}$ can be computed with
Algorithm~\ref{alg:transitionalgorithm1}.

\begin{algorithm}[tbh]
\caption{\textsc{Transitions}($\vec{V}_k, M_{min}, M_{max}, t_{min}$, $t_{max}$) : Transitions from state $\vec{V_k}$ subjected to encounters from $M_{min}$ to $M_{max}$, and tests from $t_{min}$ to $t_{max}$.}\small
\label{alg:transitionalgorithm1}
\scalebox{0.9}{
\begin{minipage}{1.2\textwidth}
\begin{algorithmic}[1]
\REQUIRE $\vec{V}_k = \left[S_k, A_k, I_k, R_k, O_k, D_k, Q_k, Ra_k \right]$, $M_{min}$, $M_{max}$, $t_{min}$, $t_{max}$
\ENSURE $Trans$
\STATE $Trans=\emptyset$
\FOR{$m = M_{min}$ \TO $M_{max}$}
\FOR{$t = t_{min}$ \TO $t_{max}$}
\STATE $Trans = Trans \cup \textsc{Transitions}(\vec{V}_k,m,t)$
\IENDFOR{}
\IENDFOR{}
\RETURN $Trans$
\end{algorithmic}
\end{minipage}
}
\end{algorithm}

\section{Additional experimental evaluation}
\label{sec:addexps}

Results for the experimental evaluation considering the reduced model
where there are not quarantined and there is no possibility for
asymptomatic to recover.

\begin{table}[tbh]
  \begin{center}
    \scalebox{0.9}{
\begin{tabular}{ |m{0.7cm} |m{0.7cm}| m{1cm}| m{1cm} | m{2.3cm}| m{2.3cm} |m{2.3cm} | }
  \hline
    $N$  & $C$ & $P_{\textit{min}}$ & $P_{\textit{max}}$ & Transition matrix creation (s) & STORM model creation (s) & STORM model checking (s) \\
  \hline
   5 & 1 & 0.376 & 0.672 & 0.012 & 0.002 & 0.002 \\

   10 & 1 & 0.423 & 0.872 & 0.933 & 0.421 & 0.236 \\

   15 & 1 & 0.442 & 0.931 & 17.846 & 13.734 & 7.200 \\

   20 & 1 & 0.452 & 0.948 & 174.414 & 189.355 & 117.356 \\

   25 & 1 & 0.459 & 0.953 & 1158.304 & 1569.474 & 2231.289 \\

   30 & 1 &  &  &  & memout &  \\

   10 & 2 & 0.036 & 0.275 & 1.437 & 0.805 & 0.247 \\

   15 & 3 & 0.003 & 0.110 & 41.746 & 51.842 & 7.114 \\

   20 & 4 & 0.0004 & 0.046 & 604.500 & 1314.256 & 94.942 \\

   25 & 5 &  &  &  & memout &  \\

  \hline
\end{tabular}
}
\end{center}
\caption{\label{tab:results_mdp_ssv_cluster} Experiments using the reduced model, no policy ($\mathcal{P}_{-1}$).}
\end{table}

\begin{table}[ht!]
\begin{center}
  \scalebox{0.82}{
    \begin{tabular}{ | m{1.6cm} | m{0.9cm}| m{0.7cm} | m{0.7cm} | m{1cm} | m{2.3cm}| m{2.1cm} |m{2.1cm} | }
  \hline
  Parameters & Policy & $N$  & $C$ & $P$ & Transition matrix creation (s) & STORM model creation (s) & STORM model checking (s) \\
  \hline

    ConstLow & $\mathcal{P}_{0}$ & 5 & 1 & 0.565 & 0.013 & 0.002 & 0.0004 \\

    ConstLow & $\mathcal{P}_{0}$ & 10 & 1 & 0.720 & 0.938 & 0.151 & 0.009 \\

    ConstLow & $\mathcal{P}_{0}$ & 15 & 1 & 0.775 & 17.864 & 2.452 & 0.160 \\

    ConstLow & $\mathcal{P}_{0}$ & 20 & 1 & 0.796 & 185.085 & 26.213 & 1.165 \\

    ConstLow & $\mathcal{P}_{0}$ & 25 & 1 & 0.806 & 1316.994 & 134.697 & 5.518 \\

    ConstLow & $\mathcal{P}_{0}$ & 30 & 1 & & memout & &  \\

    ConstLow & $\mathcal{P}_{0}$ & 10 & 2 & 0.176 & 1.434 & 0.247 & 0.018 \\

    ConstLow & $\mathcal{P}_{0}$ & 15 & 3  & 0.056 & 41.832 & 6.082 & 0.467 \\

    ConstLow & $\mathcal{P}_{0}$ & 20 & 4  & 0.019 & 600.543 & 76.463 & 4.916 \\

    ConstLow & $\mathcal{P}_{0}$& 25 & 5  &  &  & timeout &  \\

    \hline

    ConstHigh & $\mathcal{P}_{0}$ & 5 & 1  & 0.565 & 0.013 & 0.002 & 0.0005 \\

    ConstHigh & $\mathcal{P}_{0}$ & 10 & 1  & 0.720 & 0.947 & 0.153 & 0.009 \\

    ConstHigh & $\mathcal{P}_{0}$ & 15 & 1  & 0.775 & 18.076 & 2.496 & 0.151 \\

    ConstHigh & $\mathcal{P}_{0}$ & 20 & 1  & 0.796 & 183.782 & 22.601 & 1.153 \\

    ConstHigh & $\mathcal{P}_{0}$ & 25 & 1  & 0.806 & 1223.294 & 141.436 & 5.652 \\

    ConstHigh & $\mathcal{P}_{0}$ & 10 & 2 & 0.176 & 1.447 & 0.239 & 0.018  \\

    ConstHigh & $\mathcal{P}_{0}$ & 15 & 3 & 0.056 & 41.967 & 6.116 & 0.483 \\

    ConstHigh & $\mathcal{P}_{0}$ & 20 & 4 & 0.019 & 606.336 & 75.643 & 4.953 \\

    ConstHigh & $\mathcal{P}_{0}$ & 25 & 5 &  & timeout &  &  \\

    \hline

    Adapt1 & $\mathcal{P}_{1}$ & 5 & 1  & 0.565 & 0.013 & 0.002 & 0.0004 \\

    Adapt1 & $\mathcal{P}_{1}$ & 10 & 1  & 0.720 & 0.947 & 0.154 & 0.009 \\

    Adapt1 & $\mathcal{P}_{1}$ & 15 & 1  & 0.775 & 18.447 & 2.576 & 0.158 \\

    Adapt1 & $\mathcal{P}_{1}$ & 20 & 1  & 0.796 & 183.212 & 22.466 & 1.170 \\

    Adapt1 & $\mathcal{P}_{1}$ & 25 & 1  & 0.806 & 1221.046 & 134.130 & 5.702 \\


    Adapt1 & $\mathcal{P}_{1}$ & 10 & 2  & 0.176 & 1.455 & 0.237 & 0.018 \\

    Adapt1 & $\mathcal{P}_{1}$ & 15 & 3  & 0.056 & 44.092 & 6.341 & 0.487 \\

    Adapt1 & $\mathcal{P}_{1}$ & 20 & 4  & 0.019 & 646.790 & 82.957 & 5.172 \\

    Adapt1 & $\mathcal{P}_{1}$ & 25 & 5  &  & timeout &  &  \\

    \hline

    Adapt1 & $\mathcal{P}_{2}$ & 5 & 1  & 0.565 & 0.013 & 0.002 & 0.0004 \\

    Adapt1 & $\mathcal{P}_{2}$ & 10 & 1  & 0.720 & 0.947 & 0.154 & 0.009 \\

    Adapt1 & $\mathcal{P}_{2}$ & 15 & 1  & 0.775 & 18.955 & 2.567 & 0.157 \\

    Adapt1 & $\mathcal{P}_{2}$ & 20 & 1  & 0.796 & 183.792 & 23.435 & 1.139 \\

    Adapt1 & $\mathcal{P}_{2}$ & 25 & 1  & 0.806 & 1299.858 & 138.476 & 5.565 \\


    Adapt1 & $\mathcal{P}_{2}$ & 10 & 2  & 0.176 & 1.451 & 0.240 & 0.018 \\

    Adapt1 & $\mathcal{P}_{2}$ & 15 & 3  & 0.056 & 44.013 & 6.304 & 0.486 \\

    Adapt1 & $\mathcal{P}_{2}$ & 20 & 4  & 0.019 & 636.757 & 77.979 & 5.105 \\

    Adapt1 & $\mathcal{P}_{2}$ & 25 & 5  &  & timeout &  &  \\

    \hline

    Adapt2 & $\mathcal{P}_{1}$ & 5 & 1  & 0.367 & 0.014 & 0.003 & 0.0006 \\

    Adapt2 & $\mathcal{P}_{1}$ & 10 & 1  & 0.514 & 1.043 & 0.183 & 0.013 \\

    Adapt2 & $\mathcal{P}_{1}$ & 15 & 1  & 0.596 & 19.987 & 2.971 & 0.192 \\

    Adapt2 & $\mathcal{P}_{1}$ & 20 &  1 & 0.643 & 193.802 & 28.890 & 1.536 \\

    Adapt2 & $\mathcal{P}_{1}$ & 25 &  1 & 0.671 & 1264.983 & 164.665 & 8.181 \\


    Adapt2 & $\mathcal{P}_{1}$ & 10 & 2  & 0.082 & 1.659 & 0.311 & 0.027 \\

    Adapt2 & $\mathcal{P}_{1}$ & 15 &  3 & 0.020 & 48.194 & 8.473 & 0.664 \\

    Adapt2 & $\mathcal{P}_{1}$ & 20 &  4 & 0.005 & 699.542 & 122.907 & 7.225 \\

    Adapt2 & $\mathcal{P}_{1}$ & 25 &  5 &  & timeout &  &  \\

    \hline

    Adapt2 & $\mathcal{P}_{2}$ & 5 & 1  & 0.367 & 0.014 & 0.003 & 0.0005 \\

    Adapt2 & $\mathcal{P}_{2}$ & 10 & 1  & 0.514 & 1.065 & 0.183 & 0.012 \\

    Adapt2 & $\mathcal{P}_{2}$ & 15 & 1  & 0.596 & 20.024 & 2.970 & 0.192 \\

    Adapt2 & $\mathcal{P}_{2}$ & 20 & 1  & 0.643 & 193.420 & 27.144 & 1.416 \\

    Adapt2 & $\mathcal{P}_{2}$ & 25 & 1  & 0.671 & 1286.592 & 165.563 & 7.132 \\


    Adapt2 & $\mathcal{P}_{2}$ & 10 & 2  & 0.082 & 1.643 & 0.309 & 0.026 \\

    Adapt2 & $\mathcal{P}_{2}$ & 15 & 3  & 0.020 & 48.357 & 8.465 & 0.6662   \\

    Adapt2 & $\mathcal{P}_{2}$ & 20 & 4  & 0.005 & 690.604 & 106.317 & 7.093 \\

    Adapt2 & $\mathcal{P}_{2}$ & 25 & 5  &  & memout &  &  \\

  \hline
\end{tabular}
}
\end{center}
\caption{\label{tab:results_dtmc_ssv_cluster} Experiments using the reduced model, policies $\mathcal{P}_{0}$, $\mathcal{P}_{1}$ and $\mathcal{P}_{2}$.}
\end{table}